# Non-Destructive Measurement of *in-operando* Lithium Concentration in Batteries via X-Ray Compton Scattering

K. Suzuki,[1,a)] B. Barbiellini,[2] Y. Orikasa,[3] S. Kaprzyk,[2,4] M. Itou,[5] K. Yamamoto,[3] Yung Jui Wang,[2,6] H. Hafiz,[2] Y. Uchimoto,[3] A. Bansil,[2] Y. Sakurai,[5] and H. Sakurai[1]

[1]*Faculty of Science and Technology, Gunma University, Kiryu, Gunma 376-8515, Japan*

[2]*Department of Physics, Northeastern University, Boston, Massachusetts 02115, USA*

[3]*Graduate School of Human and Environmental Studies, Kyoto University, Sakyo-ku, Kyoto 606-8501, Japan*

[4]*Faculty of Physics and Applied Computer Science, AGH University of Science and Technology, aleja Mickiewicza 30, Krakow 30-059, Poland*

[5]*Japan Synchrotron Radiation Research Institute (JASRI), SPring-8, Sayo, Hyogo 679-5198, Japan*

[6]*Advanced Light Source, Lawrence Berkeley National Laboratory, Berkeley, California 94720, USA*

Non-destructive determination of lithium distribution in a working battery is key for addressing both efficiency and safety issues. Although various techniques have been developed to map the lithium distribution in electrodes, these methods are mostly applicable to test cells. Here we propose the use of high-energy x-ray Compton scattering spectroscopy to measure the local lithium concentration in closed electrochemical cells. A combination of experimental measurements and parallel first-principles computations is used to show that the shape parameter $S$ of the Compton profile is linearly proportional to lithium concentration and thus provides a viable descriptor for this important quantity. The merits and applicability of our method are demonstrated with illustrative examples of $Li_xMn_2O_4$ cathodes and a working commercial lithium coin battery CR2032.

**I. INTRODUCTION**

Advantages of electrochemical propulsion of electric vehicles and the need for large-scale energy storage in the grid are motivating the intense current effort toward the development of larger batteries with higher energy density and more stringent safety requirements. One of the challenges in battery engineering is to monitor inhomogeneous lithium distributions in the electrodes. *In-situ* and *in-operando* observations are needed for this purpose since electrochemical states in batteries are unstable. Many attempts have been made to observe lithium distributions by techniques such as x-ray absorption near-edge structure (XANES),[1-3] nuclear magnetic resonance (NMR),[4] X-ray diffraction,[5-8] neutron diffraction,[9-13]

---

[a)] Electronic mail: kosuzuki@gunma-u.ac.jp

particle induced γ-ray/x-ray emission (PIGE/PIXE),[14] Raman micro-spectroscopy,[15] and hard x-ray photoelectron spectroscopy (HX-PES).[16] Because of the short probe depth and/or low spatial resolution, however, the application of these techniques is limited to specifically designed test cells and small batteries. In this paper, we propose a more general method for determining lithium concentration in battery materials by using high-energy x-ray Compton scattering spectroscopy. In particular, we present a proof-of-the-concept study with the positive electrode materials $Li_xMn_2O_4$ as well as the commercial coin battery CR2032. Our technique yields bulk observations on working batteries since the x-ray photons reaching about 100 keV were used, and these photons can penetrate deep into materials, including closed electrochemical cells.

X-ray Compton scattering spectroscopy has been widely used to study the bulk electronic structure of materials.[17-30] Within the impulse approximation,[31,32] the energy spectra, hereafter called Compton profiles are given by

$$\frac{d^2\sigma}{d\Omega dE_2} = F \cdot J(p_z), \qquad (1)$$

where $J(p_z)$ is the Compton profile. The explicit form of the function $F$ is given by Ribberfors.[33] The Compton profile is related to the ground-state electron momentum density $\rho(\mathbf{p})$:

$$J(p_z) = \iint \rho(\mathbf{p}) dp_x dp_y, \qquad (2)$$

where $\mathbf{p} = (p_x, p_y, p_z)$ is electron momentum and $p_z$ is taken to lie along the direction of the scattering vector. The momentum density can be expressed as[34,35]

$$\rho(\mathbf{p}) = \sum_j n_j \left| \int \Psi_j(\mathbf{r}) \exp(-i\mathbf{p} \cdot \mathbf{r}) dr \right|^2, \qquad (3)$$

where $\Psi_j(\mathbf{r})$ is the wave function of electron in the $j$-state, and $n_j$ is the corresponding electron occupation. The index $j$ runs over all constituent atoms and orbitals. The details of composition of the measured material are thus reflected in its Compton profile. We will introduce below a shape parameter $S$ to characterize the Compton profile, and show that this parameter in electrode materials is proportional to the lithium concentration $x$, providing a viable descriptor for this important quantity.

**II. EXPERIMENTAL**

The present Compton scattering experiments were performed on the BL08W beamline at the SPring-8 synchrotron facility. The experimental Compton profiles from Li$_x$Mn$_2$O$_4$ were measured with a Cauchois-type x-ray spectrometer.[36-38] The incident x-ray energy was 115 keV and the scattering angle was 165 degrees. Size of the incident x-ray beam was 1.8 mm in height and 2.0 mm in width. Measurements were performed under vacuum conditions at room temperature. The overall momentum resolution was 0.1 atomic unit (a.u.). Polycrystalline Li$_x$Mn$_2$O$_4$ samples ($x$ = 0.5, 1.1, 1.2, 1.8, 1.9, 2.0, 2.1 and 3.3) were prepared by extracting or inserting lithium chemically. Pellets of these samples, 10 mm in diameter and 2 mm in thickness, were produced with a cold isostatic press for Compton scattering experiments. X-ray powder diffraction analysis was used to confirm the presence of a single spinel structure for 0.5≤ $x$ ≤1.2, and the co-existence of spinel and tetragonal structures for 1.2< $x$ ≤3.3. Lithium concentration was independently measured in all samples via the inductively coupled plasma (ICP) technique.

Details of the experimental setup for the coin battery considered are given in Ref. 39. Briefly, the energy of incident x-rays was 115 keV, and the scattering angle was 90 degrees. The overall momentum resolution was approximately 0.5 a.u., since a Ge solid-state detector was used instead of a Cauchois-type spectrometer. Size of the region probed in the coin battery is 0.1 × 0.5 × 0.5 mm$^3$. The battery is composed of a MnO$_2$ positive electrode, a negative Li electrode and a separator of olefin-based nonwoven fabric, with thicknesses of 1800 μm, 600 μm and 100 μm, respectively. The battery was discharged under a constant current (5.5 mA) for 15.75 h.

**III. RESULTS and DISCUSSION**

**A. Demonstration of the present *S*-parameter method with the example of Li$_x$Mn$_2$O$_4$**

Figure 1(a) shows the measured Compton profiles for Li$_x$Mn$_2$O$_4$ with $x$ = 0.5, 1.1 and 2.0. The areas under the profiles have been normalized to the same value. The peak height at 80 keV, for example, which is seen to increase with increasing lithium concentration, demonstrates the sensitivity of the profile to lithium concentration. The x-ray energy on the horizontal axis has been converted into electron momentum, $p_z$, by using the standard relation[31]

$$\frac{p_z}{mc} \cong \frac{E_2 - E_1 + (E_2 E_1/mc^2)(1-\cos\theta)}{\sqrt{E_1^2 + E_2^2 - 2E_1 E_2 \cos\theta}} \quad . \tag{4}$$

Here, $E_1$ and $E_2$ are energies of the incident and Compton scattered x-rays, respectively, $m$ is the electron mass, $c$ is the speed of light, and $\theta$ is the scattering angle. In Fig. 1(a), 74 keV, 80 keV and 86 keV correspond to electron momenta of -6 a.u., 0

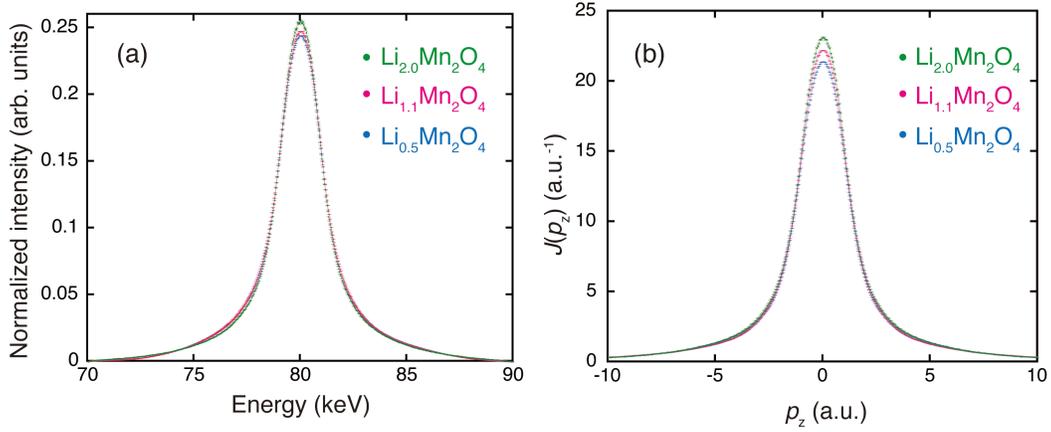

FIG. 1. (a) Energy spectra of Compton scattered x-rays and (b) Compton profiles of polycrystalline $Li_xMn_2O_4$ ($x$=0.5, 1.1 and 2.0). Blue, red and green dots correspond to the profiles of $Li_{0.5}Mn_2O_4$, $Li_{1.1}Mn_2O_4$ and $Li_{2.0}Mn_2O_4$, respectively. Compton profiles here include both the valence and core electrons.

a.u. and +6 a.u., respectively. After performing corrections for scattering cross section, x-ray absorption and multiple scattering, one obtains the profiles shown in Fig. 1(b), where large changes as a function of lithium concentration are observed around $p_z$ = 0 a.u.

The calculated atomic Compton profiles[40] of lithium ($J_{Li}$), manganese ($J_{Mn}$) and oxygen ($J_O$) atoms, and $LiMn_2O_4$ are shown in Fig. 2(a). The profile of $LiMn_2O_4$ ($J_{LMO}$) is approximated by the sum of the lithium, manganese and oxygen profiles weighted by their compositions. Both the valence and core electrons are included in these atomic profile computations. Figure 2(b) shows the valence electron contributions, $J^{val}_{Mn}$ and $J^{val}_{O}$, for manganese and oxygen atoms, where $J^{val}_{Mn}$ includes 3$d$ and 4$s$ electrons, and $J^{val}_{O}$ includes 2$s$ and 2$p$ electrons. Among these, the lithium atom ($J_{Li}$) shows the narrowest momentum distribution, which is spread within $|p_z|$ = 2.5 a.u., while both $J^{val}_{Mn}$ and $J^{val}_{O}$ extend to $|p_z|$ = 6 a.u. Notably, as expected, the core electron contributions of manganese and oxygen atoms are more extended. The valence electron states change with lithium insertion or extraction. More realistic models based on molecular orbitals and solid-state wave functions (see below) show that the variations in the Compton profile mostly occur at momenta less than $|p_z|$ = 6 a.u; this is also reflected in the lithium concentration dependence shown in Fig. 1.

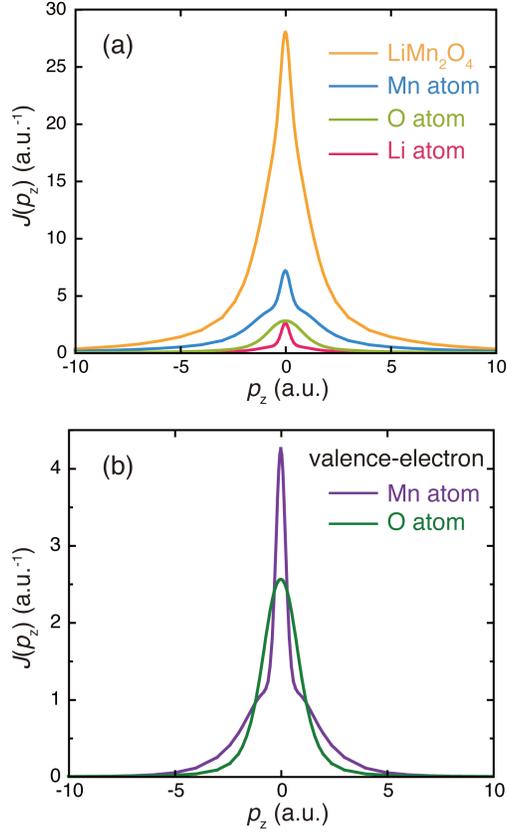

FIG. 2. (a) Compton profiles of lithium (red line), manganese (blue line) and oxygen (green line) atoms computed within the Hartree-Fock scheme of Ref. 40. The Compton profile for $LiMn_2O_4$ (orange line) is obtained by taking a weighted sum of atomic profiles of lithium, manganese and oxygen. (b) Valence-electron Compton profiles of manganese (purple line) and oxygen (green line) atoms.

The modification of the Compton profile with lithium insertion can be characterized by the *S*-parameter, which gives the relative area over a suitably defined low momentum range compared to the total area under the profile. Doppler broadening positron annihilation spectroscopy measurements[41] have already used such an *S*-parameter to characterize Li-battery cathode materials. However, positron experiments cannot be performed *in-situ* or *in-operando* because positron is a charged particle, and it cannot penetrate a closed electrochemical cell. Moreover, the interpretation of the data is complicated by positron density distribution effects in the material.[42] More specifically, we define the *S*-parameter here as the ratio

$$S = \frac{S_L}{S_H}, \qquad (5)$$

where $S_L$ and $S_H$ are the areas under the Compton profile covering the low and high momentum regions,

$$S_L = \int_{-d}^{d} J(p_z) dp_z ,\qquad(6)$$

$$S_H = \int_{-10}^{-d} J(p_z) dp_z + \int_{d}^{10} J(p_z) dp_z .\qquad(7)$$

The parameter $d$ defines the range of the low momentum region. By decomposing $S_L$ and $S_H$ into their atomic contributions ($S_{L\_Li}$, $S_{H\_Li}$, $S_{L\_O}$, $S_{H\_O}$, $S_{L\_Mn}$, and $S_{H\_Mn}$) and neglecting small interference effects, the $S$-parameter can be written as

$$S = \frac{S_L}{S_H} = \frac{xS_{L\_Li} + 2S_{L\_Mn} + 4S_{L\_O}}{xS_{H\_Li} + 2S_{H\_Mn} + 4S_{H\_O}} ,\qquad(8)$$

where the coefficients $x$, 2 and 4 reflect the stoichiometry of $Li_xMn_2O_4$. We choose the range $d = 6$ a.u. since $S_{H\_Li}$ is negligible beyond this cut-off. Thus, Eq. (8) leads to an $S$-parameter which is linearly dependent on the lithium concentration $x$. As pointed out already, lithium contribution is negligible for momenta larger than 2.5 (a.u.); it is important, however, to choose a larger value of $d$ (here 6 a.u.), so that manganese and oxygen valence states modified by lithium insertion/extraction do not give any contribution in the high momentum area of Eq. (7).

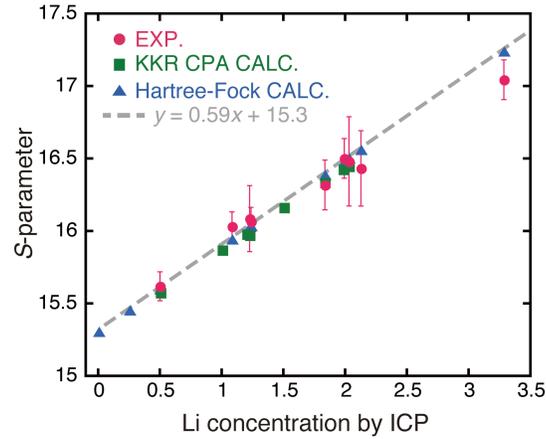

FIG. 3. Experimental $S$-parameter (red solid circles) for $Li_xMn_2O_4$ as a function of the lithium concentration, together with the corresponding theoretical values based on the atomic model (blue solid triangles) and KKR CPA first-principles computations (green solid squares). The dotted line shows the calibration curve. The $S$-parameter is seen to follow a linear variation with lithium concentration. The lithium concentration shown on the horizontal axis was determined by independent ICP measurements for various specimens used in the experiments.

Figure 3 shows a linear relationship between the $S$-parameter and the lithium concentration $x$ extracted from our measurements. In order to gain insight into this linear behavior, we have evaluated the $S$-parameter for $Li_xMn_2O_4$ for $x = 0.5$, 1.0, 1.2, 1.5, 1.8 and 2.0 using first-principles Korringa-Kohn-Rostoker coherent-potential-approximation (KKR CPA) calculations within the framework of the local spin-density approximation (LSDA).[43-45] Although the LSDA KKR CPA produces a metallic state, an energy gap can be obtained by using a Hubbard U correction. We emphasize, however, that the Compton profile and S-parameter are not sensitive to the presence of band gaps in the underlying electronic spectrum. This insensitivity reflects effects of the summation over occupied states in the definition of the momentum density in Eq. 3, and the further integration over momenta involved in computing the S-parameter. As a result, effects of redistribution of spectral weights due to band gaps in the spectrum tend to be washed out in the Compton profile and the S-parameter. Notably, the LSDA based computed Compton profiles as a function of lithium concentration are in excellent agreement with the corresponding experimental spectra.[46] Fig. 3 further shows that the S-parameter values obtained via the KKR CPA calculations as well as those obtained from the Hartree-Fock atomic model[40] justify the linear fit

$$S = 0.59x + 15.3 \ . \tag{9}$$

Therefore, such a linear fit can be used reliably to find the lithium concentration in samples where the lithium concentration is not known. Figure 4 shows the estimated statistical errors associated with the S-parameter as a function of the number of integrated x-ray photon counts $I_s$. For $d = 6$ a.u., the error, $\Delta S$, is given by the expression

$$\Delta S = 1.9 \times 10^3 I_S^{-0.5} \ . \tag{10}$$

Thus, smaller values of $d$ reduce the error. However, as discussed above, as $d$ becomes smaller, the S-parameter begins to depart from its linear behavior as a function of lithium concentration.

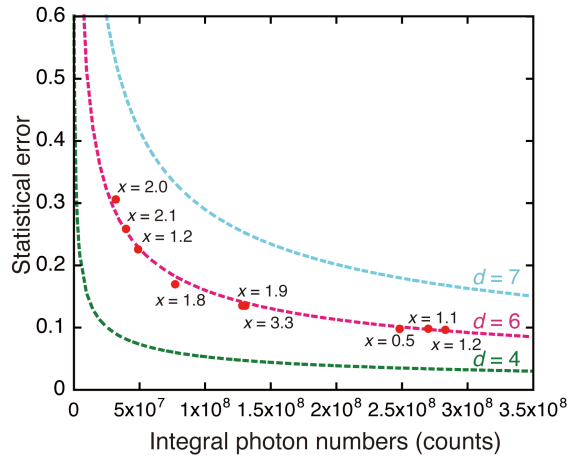

FIG. 4. Relationship between the statistical error of the S-parameter and the integrated photon number for various momentum ranges $d$: 4 (a.u.) (green dotted line), 6 (a.u.) (red dotted line) and 7 (a.u.) (blue dotted line). $x$ denotes lithium concentration.

It is interesting that Eq. 9 yields an increase in the value of the *S*-parameter with increasing *x*. This trend reflects an increase in the occupation of delocalized valence electron orbitals in the cathode, which contribute mostly to the low momentum region of the electron momentum density in Eq. (3), and lead to an increase in the *S*-parameter with increasing lithium content. By using highly precise first-principles calculations, it is possible to characterize the nature of the redox orbitals involved in the lithiation process in lithium battery cathodes as recently demonstrated by Suzuki *et al.*[46]

Intensity of Compton scattered x-rays has been used for densitometry[47] and imaging,[48,49] where the intensity scales linearly with electron density in the volume probed when x-ray absorption along both the incident and scattered x-ray beam-paths is constant. This method has been used for light-elements or low-density materials where x-ray absorption effects are small.[50,51] Recently, it has been applied to a working coin battery and the intensity variations due to lithium migration have been observed in the positive electrode.[39] However, this Compton intensity based method is not appropriate for a larger battery, since effects of x-ray absorption due to lithium redistribution become important. In contrast, the present *S*-parameter characterization has the advantage of being less sensitive to x-ray absorption effects in the target sample.

**B. Application of *S*-parameter method to a working CR2032 coin battery**

We now turn to discuss the application of our *S*-parameter characterization to the commercial lithium coin battery CR2032 under discharge conditions. By scanning the incident x-ray beam along the vertical direction of the battery, we have obtained the *S*-parameter in this case as a function of the internal height and discharge time. In order to estimate the lithium concentration through values of the *S*-parameter, we again used the momentum range of $d = 6$ a.u. For the coin battery, the *S*-parameter was normalized to its value at 0 h, which can be regarded as the value of the *S*-parameter for $MnO_2$. Although the electrodes in commercial batteries also contain an organic binder and a conductive carbon auxiliary agent, we have neglected these contributions in deducing the lithium concentration; we have, however, checked that the associated regions are small enough in the battery to be neglected.

In Fig. 5(a), the red region identifies the lithium negative electrode, the dark blue region is the $MnO_2$ positive electrode, and the yellow region is the olefin separator. Figure 5(b) shows the evolution of the normalized *S*-parameter at height A inside the coin battery of Fig. 5(a). The lithium concentration *x* in $Li_xMnO_2$ can be deduced immediately from *S*-parameter values as shown on the right hand side axis of Fig. 5(b). Note that the normalized *S*-parameter increases by 6% during discharge, which corresponds to the formation of $Li_xMnO_2$ in the positive electrode. The $Li_xMnO_2$ formation is also confirmed by the Compton profile difference [$\Delta J(p_z)$] between 0 h and 22 h shown in Fig. 5(c), where one can observe the appearance of a narrow distribution characteristic of lithium insertion into the spinel lithium manganite $Li_xMn_2O_4$.[46] [Note that the inner electric field of the battery is too weak to significantly affect the occupied electronic orbitals and hence the Compton profile.] Finally, our analysis for the coin battery reveals that lithium concentration in $Li_xMnO_2$ changes from 0 to 0.75 for the discharge time of about 16 hours.

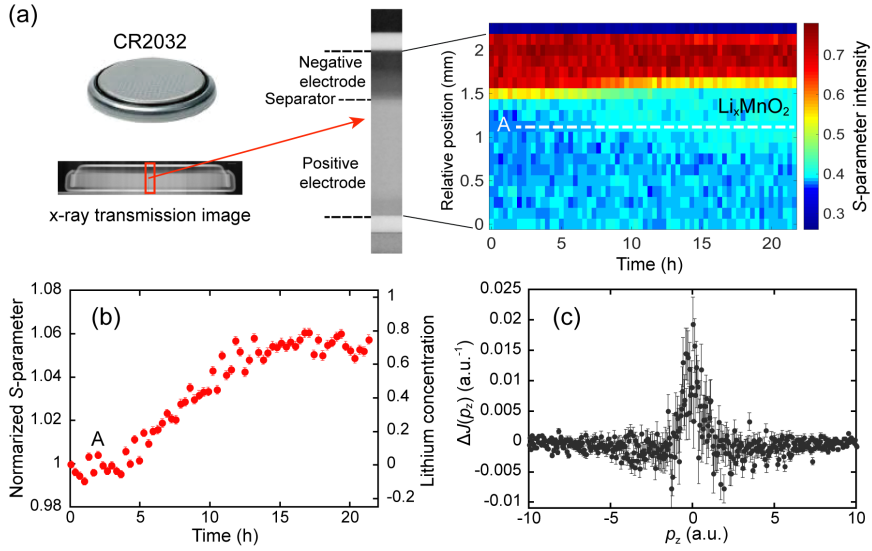

FIG. 5. (a) (Left panel) X-ray transmission image of the coin battery (CR2032) in which the region probed is highlighted by the red rectangle and, (right panel) S-parameters are shown in a false color map as a function of the internal height and discharge time. The red region corresponds to the lithium negative electrode and the dark blue region corresponds to the $MnO_2$ positive electrode. The yellow region corresponds to the olefin separator. (b) Time dependence of the normalized S-parameter along the white dotted line in Fig. 5(a). The left axis shows the normalized S-parameter while the right axis gives the lithium concentration $x$ in $Li_xMnO_2$. (c) Difference Compton profile between 0 h and 22 h.

The present method is applicable to other positive electrode materials as well as to negative electrode materials such as graphite intercalation compounds. However, a suitable range $d$ [see Eqs. (6) and (7)] must be determined in each case in order to establish the linear relationship between the S-parameter and the lithium concentration. This calibration should be checked via first-principles computations, although progress can be made via the readily available tabulated atomic data.[40] Finally, our method could be adapted to measure concentration of other ions such as sodium and magnesium used in batteries.

**IV. CONCLUSION**

We present an *in-situ/operando* x-ray Compton scattering technique for monitoring the local lithium concentration distribution in batteries. Our method is based on determining the value of the *S*-parameter, which gives the relative area under the Compton profile in a suitably defined low-momentum range compared to the total area under the profile. The *S*-parameter is shown to scale linearly with lithium concentration in both theoretical computations and experiments in $Li_xMn_2O_4$ cathode materials over a wide range of *x*-values. Results of our method for a working commercial lithium coin battery are presented to demonstrate the merits of our method as an advanced characterization technique for battery materials. Since high energy x-rays can penetrate bulk materials deeply, our technique is especially well suitable for mapping lithium distribution in large batteries, which are practically inaccessible to the currently available more conventional methods.

**ACKNOWLEDGMENTS**

This work at Gunma University, JASRI and Kyoto University was supported by Japan Science and Technology Agency. K.S. was supported by Grant-in-Aid for Young Scientists (B) from MEXT KAKENHI, Grant No. 24750065, Japan. The Compton scattering experiments were performed with approval of JASRI (proposal Nos. 2011A1869, 2011B2004, and 2012B1470). The work at Northeastern University was supported by the US Department of Energy, Office of Science, Basic Energy Sciences Grant No. DE-FG02-07ER46352, and benefited from Northeastern University's Advanced Scientific Computation Center (ASCC), and the allocation of time at the NERSC supercomputing center through DOE Grant No. DE-AC02-05CH11231. S.K. was supported by the Polish National Science Center (NCN) under Grant No. DEC-2011/02/A/ST3/00124.